\documentclass[3p,times,procedia]{elsarticle}
\usepackage{nupha_ecrc}

\volume{00}

\firstpage{1}

\journalname{Nuclear Physics A}

\runauth{}

\jid{nupha}

\jnltitlelogo{Nuclear Physics A}

\usepackage{amssymb}

\usepackage[figuresright]{rotating}

\begin{document}

\begin{frontmatter}



\dochead{XXVIIth International Conference on Ultrarelativistic Nucleus-Nucleus Collisions\\ (Quark Matter 2018)}

\title{Dynamical energy loss formalism: from describing suppression patterns to implications for future experiments}


\author{Magdalena Djordjevic$^a$, Dusan Zigic$^a$, Bojana Blagojevic$^a$, Jussi Auvinen$^a$, Igor Salom$^a$ and Marko Djordjevic$^b$}
\address{$^a$ Institute of Physics Belgrade, University of Belgrade, Serbia}
\address{$^b$ Faculty of Biology, University of Belgrade, Serbia}

\begin{abstract}
Understanding properties of Quark-Gluon Plasma requires an unbiased comparison of experimental data with theoretical predictions. To that end, we developed the dynamical energy loss formalism which, in distinction to most other methods, takes into account a realistic medium composed of dynamical scattering centers. The formalism also allows making numerical predictions for a wide number of observables with the same parameter set fixed to standard literature values. In this proceedings, we overview our recently developed DREENA-C and DREENA-B frameworks, where DREENA is a computational implementation of the dynamical energy loss formalism, and where C stands for constant temperature QCD medium, while B stands for the medium modeled by 1+1D Bjorken expansion. At constant temperature our predictions overestimate $v_2$, in contrast to other models, but consistent with simple analytical estimates. With Bjorken expansion, we have a good agreement of the predictions with both $R_{AA}$ and $v_2$ measurements. We find that introducing medium evolution has a larger effect on $v_2$ predictions, but for precision predictions it has to be taken into account in $R_{AA}$ predictions as well. Based on numerical calculations and simple analytical derivations, we also propose a new observable, which we call path length sensitive suppression ratio, for which we argue that the path length dependence can be assessed in a straightforward manner. We also argue that $Pb+Pb$ vs. $Xe+Xe$ measurements make a good system to assess the path length dependence. As an outlook, we expect that introduction of more complex medium evolution (beyond Bjorken expansion) in the dynamical energy loss formalism can provide a basis for a state of the art QGP tomography tool $–$ e.g. to jointly constrain the medium properties from the point of both high pt and low pt data.
\end{abstract}


\end{frontmatter}


\section{Introduction}
\label{}
Energy loss of high-pt particles traversing QCD medium is considered to be an excellent probe of QGP properties~\cite{QGP1,QGP2,QGP3}.
The theoretical predictions can be generated and compared with a wide range of experimental data, coming from different experiments, collision systems, collision energies, centralities, observables. This comprehensive comparison of theoretical predictions and high $p_\perp$ data, can then be used together with low $p_\perp$ theory and data to study the properties of created QCD medium~\cite{JET,Aarts:2016hap,Akiba,Brambilla}, that is, for precision QGP tomography. However, to implement this idea, it is crucial to have a reliable high $p_\perp$ parton energy loss model. With this goal in mind, during the past several years, we developed the dynamical energy loss formalism~\cite{MD_Dyn}. Contrary to the widely used approximation of static scattering centers, this model takes into account that QGP consists of dynamical (moving) partons, and that the created medium has finite size. The calculations are based on the finite temperature field theory, and generalized HTL approach. The formalism takes into account both radiative and collisional energy losses, is applicable to both light and heavy flavor, and has been recently generalized to the case of finite magnetic mass and running coupling~\cite{MD_PLB}. Most recently, we also relaxed the soft-gluon approximation within the model~\cite{BDD}. Finally, the formalism is integrated in an up-to-date numerical procedure~\cite{MD_PLB}, which contains parton production, fragmentation functions, path-length and multi-gluon fluctuations.

The model up-to-now explained a wide range of $R_{AA}$ data~\cite{MD_PLB,DBZ,MD_5TeV,MD_PRL}, with the same numerical procedure, the same parameter set, and with no fitting parameters, including explaining puzzling data and generating predictions for future experiments. This then strongly suggests that the model provides a realistic description of high $p_\perp$ parton-medium interactions. However, the model did not take into account the medium evolution, so we used it to provide predictions  only for those observables that are considered to be weakly sensitive to QGP evolution.

Therefore, our goal, which will be addressed in this proceedings, is to develop a framework which will allow systematic comparison of experimental data and theoretical predictions, obtained by the same formalism and the same parameter set. In particular, we want to develop a framework, which can systematically generate predictions for different observables (both $R_{AA}$ and $v_2$), different collision systems ($Pb+Pb$ and $Xe+Xe$), different probes (light and heavy), different collision energies and different centralities~\cite{DREENA_C,DREENA_B,DREENA_Xe}. Within this, our major goal is to introduce medium evolution in the dynamical energy loss formalism~\cite{DREENA_B}, where we start with 1+1D Bjorken expansion~\cite{BjorkenT}, and where our developments in this direction, will also be outlined in this proceedings. Finally, we also want to address an important question of how to differentiate between different energy loss models; in particular, what is appropriate observable, and what are appropriate systems, to assess energy loss path-length dependence~\cite{DREENA_Xe}. Note that only the main results are presented here; for a more detailed version, see~\cite{DREENA_C,DREENA_B,DREENA_Xe}, and references therein.

\section{Results and discussion}
As a first step towards the goals specified above, we developed DREENA-C framework~\cite{DREENA_C}, which is a fully optimized computational suppression procedure based on our dynamical energy loss formalism in constant temperature finite size QCD medium. With this framework we, for the first time, generated joint $R_{AA}$ and $v_2$ predictions within our dynamical energy loss formalism. We generated predictions for both light and heavy flavor probes, and different centrality regions in $Pb+Pb$ collisions at the LHC (see~\cite{DREENA_C} for more details). We obtained that, despite the fact that DREENA-C does not contain medium evolution (to which $v_2$ is largely sensitive), it leads to qualitatively good agreement with this data, though quantitatively, the predictions are visibly above the experimental data.

The theoretical models up-to-now, faced difficulties in jointly explaining $R_{AA}$ and $v_2$ data, i.e. lead to underprediction of $v_2$, unless new phenomena  are introduced, which is known as $v_2$ puzzle~\cite{v2Puzzle}. Having this in mind, the overestimation of $v_2$, obtained by DREENA-C, seems surprising. However, by using a simple scaling arguments, where fractional energy loss is proportional to $T^a$ and $L^b$ , and where, within our model $a, b$ are close to 1, we can straightforwardly obtain that, in constant temperature medium, $R_{AA} \approx 1-\xi T L$ and $v_2 \approx \frac{ \xi T \Delta L}{2}$, while in evolving medium they have the following expressions $R_{AA} \approx 1-\xi T L$ and $v_2 \approx \frac{\xi T \Delta L - \xi \Delta T L}{2}$ (see~\cite{DREENA_C} for more details, $\xi$ is a proportionality factor that depends on initial jet $p_\perp$)). So, it is our expectation that, within our model, the medium evolution will not significantly affect $R_{AA}$, while it will notably lower the $v_2$ predictions.

To check the reliability of these simple estimates, we developed DREENA-B framework~\cite{DREENA_B}, which is our most recent development within dynamical energy loss formalism. Here B stands for 1+1D Bjorken expansion~\cite{BjorkenT}, i.e. the medium evolution is introduced in dynamical energy loss formalism in a simple analytic way. We provided first joint $R_{AA}$ and $v_2$ predictions with dynamical energy loss formalism in expanding QCD medium, which are presented in Fig.~1 (for charged hadrons), and we observe very good joint agreement with $R_{AA}$ and $v_2$ data. We equivalently obtained the same good agreement for D mesons, and predicted non-zero $v_2$ for high $p_\perp$ B mesons.

\begin{figure}[h]
\includegraphics[scale=0.23]{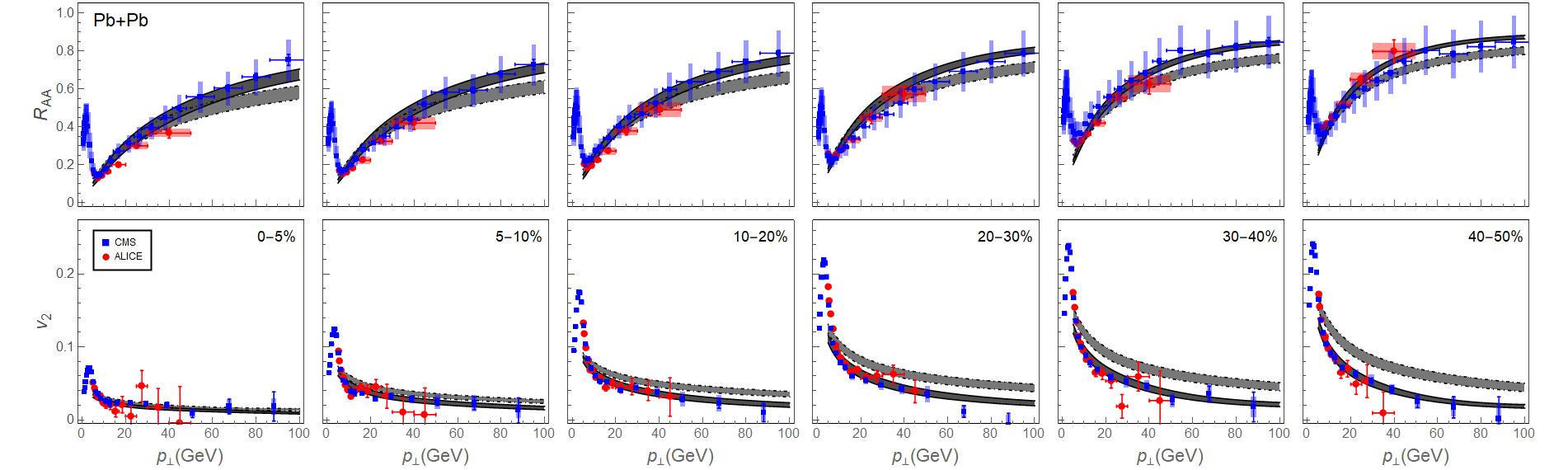}
\caption{ {\bf Joint $R_{AA}$ and $v_2$ predictions for charged hadrons in $5.02$~TeV  $Pb+Pb$ collisions.} {\it Upper panels:} Predictions for $R_{AA}$ {\it vs.} $p_\perp$ are compared with  ALICE~\cite{ALICE_CH_RAA} (red circles) and CMS~\cite{CMS_CH_RAA} (blue squares) charged hadron experimental data in $5.02$~TeV $Pb+Pb$ collisions.  {\it Lower panels:} Predictions for $v_2$ {\it vs.} $p_\perp$ are compared with ALICE~\cite{ALICE_CH_v2} (red circles) and CMS~\cite{CMS_CH_v2} (blue squares) experimental data in $5.02$~TeV $Pb+Pb$ collisions. Full and dashed curves correspond, respectively, to the predictions obtained with DREENA-B and DREENA-C frameworks. Columns 1-6 correspond, respectively, to $0-5\%$, $5-10\%$, $10-20\%$,..., $40-50\%$ centrality regions. The figure is adapted from~\cite{DREENA_C,DREENA_B} and the parameter set is specified there. }
\label{fig1}
\end{figure}

In Fig.~2, we further present predictions for $Xe+Xe$ data~\cite{DREENA_Xe}, where we note that these predictions were generated before the data became available. In this figure (see also Fig.~1), we compare DREENA-C and DREENA-B frameworks, to assess the importance of including medium evolution on $R_{AA}$ and $v_2$ observables. We see that inclusion of medium evolution has effect on both $R_{AA}$ and $v_2$ data. That is, it systematically somewhat increase $R_{AA}$, while significantly decreasing $v_2$; this observation is in agreement with our estimate provided above. Consequently, we see that this effect has large influence on $v_2$ predictions, confirming previous  arguments that $v_2$ observable is quite sensitive to medium evolution. On the other hand, this effect is rather small on $R_{AA}$, consistent with the notion that $R_{AA}$ is not very sensitive to medium evolution. However, our observation from Figs.~1 and~2 is that medium evolution effect on $R_{AA}$, though not large, should still not be neglected in precise $R_{AA}$ calculations, especially for high pt and higher centralities.
\begin{figure}[htb]
\begin{minipage}[t]{70mm}
\includegraphics[scale=0.66]{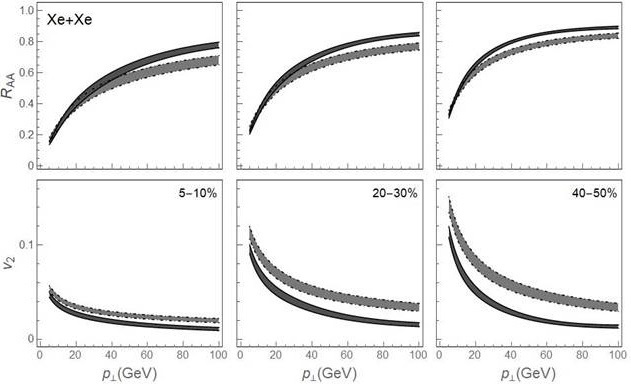}
\vspace{-0.6cm}
\caption{ {\bf Joint $R_{AA}$ and $v_2$ predictions for charged hadrons in $5.44$~TeV $Xe+Xe$ collisions.} Predictions for and $R_{AA}$ {\it vs.} $p_\perp$ and $v_2$ {\it vs.} $p_\perp$ are shown on upper and lower panels, respectively.  Columns 1-3, respectively, correspond to $5-10\%$, $20-30\%$ and $40-50\%$ centrality regions. Full and dashed curves correspond, respectively, to the predictions obtained with DREENA-B and DREENA-C frameworks.  In each panel, the upper (lower) boundary of each gray band corresponds to $\mu_M/\mu_E =0.6$ ($\mu_M/\mu_E =0.4$). The figure is adapted from~\cite{DREENA_B} and the parameter set is specified there. }
\label{fig2}
\end{minipage}
\hspace{\fill}
\begin{minipage}[t]{70mm}
\includegraphics[scale=0.43]{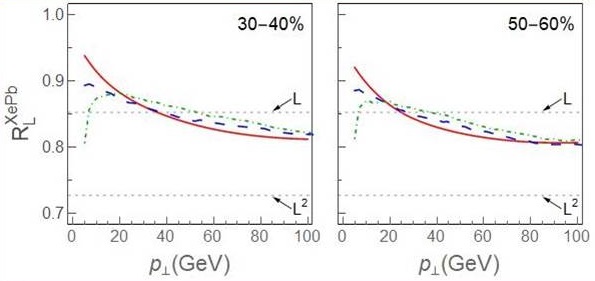}
\vspace{-0.6cm}
\caption{ {\bf Path-length sensitive suppression ratio ($R_L^{XePb}$) for light and heavy probes.} Predictions for $R_L^{XePb}$ {\it vs.} $p_\perp$ is shown for charged hadrons (full), D mesons (dashed) and B mesons (dot-dashed). First and second column, respectively, correspond to $30-40\%$ and $50-60\%$ centrality regions.  $\mu_M/\mu_E =0.4$. The figure is adapted from~\cite{DREENA_Xe} and the parameter set is specified there. }
\label{fig3}
\end{minipage}
\end{figure}

Finally, as the last topic of this proceedings, we address a question on how to differentiate between different energy loss models. With regard to this, note that path length dependence provides an excellent signature differentiating between different energy loss models, and consequently also between the underlying energy loss mechanisms. For example, some energy loss models have linear, some have quadratic path-length dependence, and the dynamical energy loss path-length dependence is between linear and quadratic, which is due to both collisional and radiative energy loss mechanisms included in the model. To address this question, we first have to answer what is an appropriate system for such a study. We argue that comparison of suppressions in $Pb+Pb$ and $Xe+Xe$ is an excellent way to study the path length dependence: From the suppression calculation perspective, almost all properties of these two systems are the same. That is, we show~\cite{DREENA_Xe} that these two systems have very similar initial momentum distributions, average temperature for each centrality region and path length distributions (up to rescaling factor $A^{1/3}$). That is, the main property differentiating the two systems is its size, i.e. rescaling factor $A^{1/3}$, which therefore makes comparison of suppressions in $Pb+Pb$ and $Xe+Xe$ collisions an excellent way to study the path length dependence.

The second question is what is appropriate observable? With regards to that, the ratio of the two $R_{AA}$ seems a natural choice, as has been proposed before.
However, in this way the path length dependence cannot be naturally extracted, as shown in~\cite{DREENA_Xe}. For example, this ratio approaches one for high $p_\perp$ and high centralities, suggesting no path length dependence, while the dynamical energy loss used to generate this figure has strong path length dependence. Also, the ratio has strong centrality dependence. That is, from this ratio, no useful information can be deduced. The reason for this is that this ratio includes a complicated relationship (see~\cite{DREENA_Xe} for more details) which depends on the initial jet energy and centrality; so extracting the path-length dependence from this observable would not be possible.

However, based on the derivation presented in~\cite{DREENA_Xe}, we propose to use the ratio of 1-$R_{AA}$ instead. From this estimate, we see that this ratio $R_L^{XePb} \equiv \frac{1-R_{XeXe}}{1-R_{PbPb}} \approx \left(\frac{A_{Xe}}{A_{Pb}}\right)^{b/3}$ has a simple dependence on only the size of the medium ($A^{1/3}$ ratio) and the path length dependence (exponent $b$). In Fig.~3 we plot this ratio, where we see that the path length dependence can be extracted from this ratio in a simple way, and moreover there is only a weak centrality dependence. Therefore, 1-$R_{AA}$ ratio seems as a natural observable, which we propose to call path-length sensitive suppression ratio.

\bigskip
{\bf Acknowledgements:}
This work is supported by the European Research Council, grant ERC-2016-COG: 725741, and by the Ministry of Science and Technological
Development of the Republic of Serbia, under project numbers ON171004, ON173052 and ON171031.

\end{document}